\documentclass[thmsa]{article}
\usepackage{sw20lart}
\input tcilatex
\QQQ{Language}{
British English
}

\begin{document}

\section{\bf INTRODUCTION}

The conceptual foundations of quantum mechanics continue to be a subject of
debate and different interpretations of the theory still contend for
acceptance. In this paper, we employ the the interpretation of quantum
mechanics due to Land\'e$[1,2]$ on spin theory and thereby to obtain a new
and more generalized treatment of spin.

In this work, we show how to derive the spin operators and eigenvectors from
probability amplitudes. We obtain generalized forms of these quantities. To
the best of our knowledge, both the method of derivation and the generalized
expressions we present are new. The method shows that spin operators and
vectors can be derived from an analysis which takes the probability
amplitudes that characterize spin measurements as being primary, in the same
way that the spherical harmonics and differential orbital angular momentum
operators might be taken as being more fundamental than column vectors and
matrices in describing orbital angular momentum. Thus the method we
introduce does not require us to study the form of orbital angular momentum
matrices and column vectors in order to construct spin vectors and operators
by analogy. At the same time, this method places the treatment of spin on an
even more analogous footing with that of orbital angular momentum than
hitherto.

This paper is constructed as follows. In Section $2$, we give a brief review
of the Land\'e approach to quantum mechanics, because it is likely to be
unfamiliar to most readers. In Section $3$ we use the Land\'e approach to
investigate the transition from wave to matrix mechanics. In Section $4$ we
specialize the results of the previous sections to the case of spin-$1/2$.
We show in Subsection $4.1$ how the matrix treatment of spin may be deduced
from an approach that starts with probability amplitudes. In Subsection $4.3$%
, we show how to obtain the Pauli spin matrices and their eigenvectors by
this approach. We derive explicit formulas for the probability amplitudes
characterizing spin-$1/2$ measurements in Subsection $4.4$. In Subsections $%
4.7$, $4.8$ and $4.9$, we obtain the most general forms of the spin
operators and give the eigenvectors of the generalized $z$ component. In
Subsection $4.10$, we show that in the appropriate limit, the generalized
operators and vectors reduce to the Pauli forms, and to those forms
heretofore considered to be the most general . In Subsection $4.11$, we
discuss the connection between our results and the standard results. We
conclude the paper after a Summary and Discussion in Section $5$.

\section{\bf REVIEW OF THE LAND\'E APPROACH}

\subsection{\bf Basic Results}

We begin by giving those results of the Land\'e approach which we shall need
for our purpose. The contents of this section will be found set out at
complete length in Refs.$1$ and $2$.

According to Land\'e, we must start by accepting that indeterminism is
inherent in nature, and therefore in quantum measurement. Therefore upon
repeated measurement, different results may follow from exactly the same
preparation of a system. Statistical theory predicts the probabilities of
obtaining the several different outcomes of any such measurement.

We may briefly characterize quantum mechanics as dealing with material
bodies composed of particles with coordinates and momenta $q$ and $p$ and
with various observable quantities $A$,$B$,$C$,.. defined as functions of
the $q$'s and $p$'s. Measurement forces the measured quantity to take one of
the eigenvalues of the operator representing the measured quantity. The
observable $A$ can manifest many possible values $A_1$, $A_2$, $A_3,...$
when it is measured.

Suppose a system is in a state characterized by the eigenvalue $A_i$. A
measurement of the quantity $B$ is then made. Any one of the values of the
eigenvalue spectrum of $B$ can result. The probabilities of obtaining the
different values of $B$ are $P(A_i;B_j).$ Conversely, if we start with the
system being in a state characterized by the eigenvalue $B_j$ and make a
measurement of $A$, then any of the values $A_1$, $A_2$, $A_3$..... can
result. The probabilities of obtaining the different values of $A$ are $%
P(B_j;A_i)$. The two sets of probabilities obey two-way symmetry. Hence

\begin{equation}
P(A_i;B_j)=P(B_j;A_i).  \label{one}
\end{equation}
The probabilities sum to unity:

\begin{equation}
\dsum\limits_jP(A_i;B_j)=\dsum\limits_iP(B_j;A_i)=1.  \label{two}
\end{equation}
To ensure reproducibility of results upon repetition of measurement, the
probabilities satisfy

\begin{equation}
P(A_i;A_j)=\delta _{ij}.  \label{three}
\end{equation}

We shall assume that both $A$ and $B$ have the same number of eigenvalues $N$%
; this can be shown to be always so$[3]$. We arrange the probabilities in a
table to obtain the matrix

\begin{equation}
\left[ P_{AB}\right] =\left( 
\begin{array}{cccc}
P(A_1;B_1) & P(A_1;B_2) & ... & P(A_1;B_N) \\ 
P(A_2;B_1) & P(A_2;B_2) & ... & P(A_2;B_N) \\ 
... & ... & ... & ... \\ 
P(A_N;B_1) & P(A_N;B_2) & ... & ...P(A_N;B_N)
\end{array}
\right) .  \label{four}
\end{equation}

Each row sums to unity, and, due to the symmetry of the probabilities,
contained in Eqn. (\ref{one}), so does each column. Since $P(A_i;B_j)\geq 0,$
we may set $\alpha ^2(A_i;B_j)=P(A_i;B_j);$ the $\alpha ^{\prime }$s are
probability amplitudes. We now have

\begin{equation}
\left[ P_{AB}\right] =[\alpha _{AB}^2]=\left( 
\begin{array}{cccc}
\alpha ^2(A_1;B_1) & \alpha ^2(A_1;B_2) & ... & \alpha ^2(A_1;B_N) \\ 
\alpha ^2(A_2;B_1) & \alpha ^2(A_2;B_2) & ... & \alpha ^2(A_2;B_N) \\ 
... & ... & ... & ... \\ 
\alpha ^2(A_N;B_1) & \alpha ^2(A_N;B_2) & ... & \alpha ^2(A_N;B_N)
\end{array}
\right) .  \label{five}
\end{equation}

As each row and each column sums to unity, this suggests that the $\alpha $%
's are direction cosines, where $A_1$, $A_2$,...,$A_N$ are $N$ mutually
perpendicular axes in $N$-dimensional space and $B_1$, $B_2$,...$B_N$ is
another set of orthogonal axes. This means that if another observable $C$
with $N$ eigenvalues $C_1$, $C_2$,...$C_M$ is introduced, the three sets of $%
\alpha $'s are connected through the relation

\begin{equation}
\alpha (A_i;C_k)=\dsum\limits_{j=1}^N\alpha (A_i;B_j)\alpha (B_j;C_k)
\label{six}
\end{equation}
which holds for direction cosines. Though we have $\alpha =\sqrt{P}$, the
most general case involves a phase, so that

\begin{equation}
\alpha (A_i;C_k)=\psi (A_i;C_k)=\sqrt{P(A_i;C_k)}e^{i\varphi }  \label{seven}
\end{equation}
where $\varphi \;$is the phase. This choice also leaves the rows and columns
of the matrix in Eqn. (\ref{five}) summing to unity. The two-way symmetry
expressed in Eqn. (\ref{one}) is now contained in the Hermiticity condition

\begin{equation}
\psi (A_i;C_k)=\psi ^{*}(C_k;A_i).  \label{eit}
\end{equation}

Eqn. (\ref{six}) now connects the probability amplitudes $\psi (A_i;C_k)$, $%
\chi (A_i;B_{_j})\;$and $\phi (B_j;C_k),\;$and takes the form

\begin{equation}
\psi (A_i;C_k)=\dsum\limits_{j=1}^N\chi (A_i;B_j)\phi (B_j;C_k).
\label{nine}
\end{equation}

We have set $\chi (A_i;B_j)=\psi (A_i;B_j)\;$and $\phi (B_j;C_k)=\psi
(B_j;C_k)\;$to avoid confusion. This expression is evidently the well-known
law of probability interference. It is the basis of the expansion of the
wave function in terms of a complete set. To see this, we omit $A_i$, the
label for the initial state and let $C_k=x$ because the final eigenvalue
spectrum is continuous. Then we set

\begin{equation}
b_j=\chi (A_i;B_j),\;\;\;\;\;\phi _j=\phi (B_j;C_k).\;  \label{ten}
\end{equation}
We then obtain the familiar expression

\begin{equation}
\psi (x)=\dsum \limits_{j=1}^Nb_j\phi _j(x).  \label{el11}
\end{equation}

\subsection{\bf Extension of Basic Results}

The following formulas are not explicitly given by Land\'e, but they are
important to what subsequently follows.

When $A=C$, Eqn. (\ref{nine}) becomes

\begin{equation}
\psi (A_i;A_k)=\dsum\limits_{j=1}^N\chi (A_i;B_j)\phi
(B_j;A_k)=\dsum\limits_{j=1}^N\chi (A_i;B_j)\chi ^{*}(B_j;A_k)=\delta _{ik}
\label{tw12}
\end{equation}
In the standard expression Eqn. (\ref{el11}), the expansion coefficients $%
b_j $ are of course, probability amplitudes. However, the extra detail in
Eqn. (\ref{nine}) shows that, far from being merely constants, these
probability amplitudes $\chi (A_i;B_j)$ have a structure. In fact, they are
completely equivalent in character to the members of the basis set $\{\phi
(B_j;C_k)\}$. Because of the fact that they are probability amplitudes, the
expansion coefficients satisfy

\begin{equation}
\dsum\limits_{j=1}^N\left| \chi (A_i;B_j)\right|
^2=\dsum\limits_{j=1}^N\left| b_j\right| ^2=1.  \label{th13}
\end{equation}

Usually, this result and the interpretation of the expansion coefficient as
probability amplitudes are justified in a more roundabout way, while the
Land\'e formalism justifies these results naturally and automatically.

Since all the functions in the expansion Eqn. (\ref{nine}) are on the same
footing, there are equivalent expressions for both $\chi (A_i;B_j)\;$and $%
\phi (B_j;C_k)$. If we multiply the basic expansion Eqn. (\ref{nine}) by $%
\chi ^{*}(A_i;B_l)\;$and sum over $i$, we obtain

\begin{equation}
\dsum\limits_{i=1}^N\chi ^{*}(A_i;B_l)\psi
(A_i;C_k)=\dsum\limits_{j=1}^N\phi (B_j;C_k)\dsum\limits_i\chi
^{*}(A_i;B_l)\chi (A_i;B_j).  \label{fo14}
\end{equation}
But in view of Eqn. (\ref{eit}) and Eqn. (\ref{tw12}), we obtain

\begin{equation}
\phi (B_l;C_k)=\sum\limits_{i=1}^N\chi (B_l;A_i)\psi (A_i;C_k).  \label{si16}
\end{equation}

The other relation is found by multiplying both sides of Eqn. (\ref{nine})
by $\phi ^{*}(B_m;C_k)\;$and summing over $k$. Due to the orthogonality of
the functions $\phi (B_j;C_k),$we obtain

\begin{equation}
\chi (A_i;B_m)=\dsum\limits_{k=1}^N\psi (A_i;C_k)\phi (C_k;B_m)  \label{se17}
\end{equation}
an expression that is better known in the form

\begin{equation}
b_m=\int \phi _m^{*}(x)\psi (x)dx.  \label{ei18}
\end{equation}

\section{\bf THE EIGENVALUE EQUATION IN MATRIX FORM}

\subsection{\bf The Differential Eigenvalue Equation and Its Interpretation
In The Land\'e Formalism}

We now proceed to an investigation of the connection between wave and matrix
mechanics. We seek to obtain the matrix eigenvalue equation from the
differential eigenvalue equation.

Though Land\'e does not explicitly treat this point, his formalism requires
the continuous variable in the differential eigenvalue equation to represent
the eigenvalues corresponding to the states being obtained upon measurement.
The eigenvalue that appears as the label of the eigenfunction in the
differential eigenvalue equation describes the state that obtains before
measurement. Thus, for example, in the single-particle time-independent
Schroedinger equation

\begin{equation}
H({\bf r})\psi _k({\bf r})=E_k\psi _k({\bf r})  \label{th33}
\end{equation}
$E_k$ represents the eigenvalue corresponding to the state that obtains
before measurement, while ${\bf r}$ represents the result of the
measurement. The eigenfunction $\psi _k({\bf r})$ is the probability
amplitude that if the system is in a state corresponding to $E_k$, the value 
${\bf r}$ is obtained upon measurement of the position. Similarly, in the
equation

\begin{equation}
-i\hbar \frac \partial {\partial x}\psi _{p_0}(x)=p_0\psi _{p_0}(x)
\label{th34}
\end{equation}
$p_0$ represents the eigenvalue for the initial state, while $x$ is the
final-state eigenvalue. Thus, in this equation, $\psi (x)$ gives the
probability amplitude for measurement of the position of the system if the
system is initially in a state characterized by the momentum $p_0$.

In the Land\'e formalism, the state that obtains before measurement is
well-defined and has been brought about by prior measurement. In a
differential eigenvalue equation, the operator acts on the variables
representing the final eigenvalue of measurement. Therefore, the
differential eigenvalue equation is possible only if the eigenfunction has a
continuous eigenvalue spectrum, so that the differential operator can act on
a continuous variable.

\subsection{\bf From The Differential to the Matrix Eigenvalue Equation}

As is well known, the differential eigenvalue equation is transformed to a
matrix eigenvalue equation in the following way. Let the eigenvalue equation
be

\begin{equation}
A(x)\psi (x)=\lambda \psi (x)  \label{th35}
\end{equation}
where $A(x)$ is the operator and $\lambda $ is the eigenvalue. We expand $%
\psi (x)$ in terms of the complete orthonormal set $\{\phi _j(x)\}$, which
has $N$ non-degenerate members:

\begin{equation}
\psi (x)=\dsum\limits_{j=1}^Nb_j\phi _j(x)  \label{th36}
\end{equation}
where $b_j$ is a constant. We substitute into Eqn. (\ref{th35});
pre-multiplying by $\phi _m^{*}$, integrating over $x$ and using the
orthogonality of the members of the set $\{\phi _j\}\;$gives

\begin{equation}
\dsum\limits_{j=1}^Nb_j\left\langle \phi _m\left| A\right| \phi
_j\right\rangle -\lambda b_m=0,\;\;\;\;\;\;m=1,2,3,......,N  \label{th38}
\end{equation}
With $A_{mj}=\left\langle \phi _m\left| A\right| \phi _j\right\rangle $, we
obtain the following system of equations if we write out these equations for
each value of $m$:

\begin{eqnarray}
\lbrack A_{11}-\lambda ]b_1-A_{12}b_2+\;...\;+A_{1N}b_N &=&0  \label{th39} \\
A_{21}b_1+[A_{22}-\lambda ]b_2+\;...\;+A_{2N}b_N &=&0  \nonumber \\
&&.  \nonumber \\
&&.  \nonumber \\
A_{N1}b_1+A_{N2}b_2+\;...\;+[A_{NN}-\lambda ]b_N &=&0  \nonumber
\end{eqnarray}

Hence the matrix eigenvalue equation is

\begin{equation}
\left( 
\begin{array}{cccc}
A_{11}-\lambda & A_{12} & ... & A_{1N} \\ 
A_{21} & A_{22}-\lambda & ... & A_{2N} \\ 
... & ... & ... & ... \\ 
... & ... & ... & ... \\ 
A_{N1} & A_{N2} & ... & A_{NN}-\lambda
\end{array}
\right) \left( 
\begin{array}{c}
b_1 \\ 
b_2 \\ 
... \\ 
... \\ 
b_N
\end{array}
\right) =0  \label{fo40}
\end{equation}
The vector

\begin{equation}
\left[ b\right] =\left( 
\begin{array}{c}
b_1 \\ 
b_2 \\ 
... \\ 
... \\ 
b_N
\end{array}
\right)  \label{fo41}
\end{equation}
is the matrix representation of the eigenfunction in the basis $\{\phi _j\}$.

The expectation value of the quantity $R(x)$ is

\begin{equation}
\left\langle R\right\rangle
=\dsum\limits_{i=1}^N\dsum\limits_{j=1}^Nb_i^{*}R_{ij}b_j=\left[ b\right]
^{\dagger }\left[ R\right] \left[ b\right]  \label{fo42}
\end{equation}
where

\begin{equation}
\left[ R\right] =\left( 
\begin{array}{cccc}
R_{11} & R_{12} & ... & R_{1N} \\ 
R_{21} & R_{22} & ... & R_{2N} \\ 
... & ... & ... & ... \\ 
R_{N1} & R_{N2} & ... & R_{NN}
\end{array}
\right)  \label{fo43}
\end{equation}
and

\begin{equation}
R_{ij}=\left\langle \phi _i\left| R(x)\right| \phi _j\right\rangle .
\label{fo44}
\end{equation}

\subsection{\bf The Matrix Eigenvalue Equation in the Land\'e Formalism}

We now recast this procedure in terms of the Land\'e formalism. We start
again with the differential eigenvalue equation: we write this out with as
much information as possible:

\begin{equation}
A(x)\psi (\lambda _k;x)=\lambda _k\psi (\lambda _k;x).  \label{fo45}
\end{equation}
Here $x$ is a general quantity which represents the final eigenvalue, whose
spectrum is continuous. We use Eqn. (\ref{nine}) to express $\psi (\lambda
_k,x)\;$as an expansion:

\begin{equation}
\psi (\lambda _k;x)=\dsum\limits_{j=1}^N\chi (\lambda _k;B_j)\phi (B_j;x).
\label{fo46}
\end{equation}
We substitute this in the eigenvalue Eqn. (\ref{fo45}). Then we pre-multiply
by $\phi ^{*}(B_m,x)\;$and integrate over $x$ to obtain

\begin{equation}
\dsum\limits_{j=1}^N\int \phi ^{*}(B_m;x)A(x)\phi (B_j;x)dx\chi (\lambda
_k;B_j)=\dsum\limits_{j=1}^N\lambda _k\int \phi ^{*}(B_m;x)\phi
(B_j;x)dx\chi (\lambda _k;B_j).  \label{fo48}
\end{equation}
Due to the orthogonality of the probability amplitudes $\phi $, we obtain

\begin{equation}
\dsum\limits_{j=1}^NA_{mj}\chi (\lambda _k;B_j)-\lambda _k\chi (\lambda
_k;B_j)=0,\;\;\;\text{for }m=1,2,3,.....N  \label{fo49}
\end{equation}
where

\begin{equation}
A_{mj}=\left\langle \phi (B_m;x)\left| A(x)\right| \phi (B_j;x)\right\rangle
.  \label{fi50}
\end{equation}
By writing down the equations for each value of $m$, we are led to the
matrix eigenvalue equation

\begin{equation}
\left( 
\begin{array}{cccc}
A_{11}-\lambda _k & A_{12} & ... & A_{1N} \\ 
A_{21} & A_{22}-\lambda _k & ... & A_{2N} \\ 
... & ... & ... & ... \\ 
... & ... & ... & ... \\ 
A_{N1} & A_{N2} & ... & A_{NN}-\lambda _k
\end{array}
\right) \left( 
\begin{array}{c}
\chi (\lambda _k,B_1) \\ 
\chi (\lambda _k,B_2) \\ 
... \\ 
... \\ 
\chi (\lambda _k,B_N)
\end{array}
\right) =0  \label{fi51}
\end{equation}
If we denote the eigenvector by 
\begin{equation}
\lbrack \xi (\lambda _k)]=\left( 
\begin{array}{c}
\chi (\lambda _k,B_1) \\ 
\chi (\lambda _k,B_2) \\ 
... \\ 
... \\ 
\chi (\lambda _k,B_N)
\end{array}
\right) ,  \label{fi51a}
\end{equation}
then we obtain the matrix eigenvalue equation

\begin{equation}
\lbrack A][\xi (\lambda _k)]=\lambda _k[\xi (\lambda _k)].  \label{fi51b}
\end{equation}

The expectation value of the quantity $R(x)$ is given by

\begin{equation}
\left\langle R\right\rangle =\dsum\limits_{i=1}^N\dsum\limits_{j=1}^N\chi
^{*}(\lambda _k;B_i)R_{ij}\chi (\lambda _k;B_j)=[\psi (\lambda _k)]^{\dagger
}\left[ R\right] [\psi (\lambda _k)]  \label{fi52}
\end{equation}
where

\begin{equation}
R_{ij}=\left\langle \phi (B_i;x)\left| R(x)\right| \phi (B_j;x)\right\rangle
\label{fi52a}
\end{equation}

\begin{equation}
\lbrack \psi (\lambda _k)]=\left( 
\begin{array}{c}
\chi (\lambda _k;,B_1) \\ 
\chi (\lambda _k;B_2) \\ 
... \\ 
... \\ 
\chi (\lambda _k;B_N)
\end{array}
\right)  \label{fi53}
\end{equation}
and

\begin{equation}
\left[ R\right] =\left( 
\begin{array}{cccc}
R_{11} & R_{12} & ... & R_{1N} \\ 
R_{21} & R_{22} & ... & R_{2N} \\ 
... & ... & ... & ... \\ 
... & ... & ... & ... \\ 
R_{N1} & R_{N2} & ... & R_{NN}
\end{array}
\right) .  \label{fi53a}
\end{equation}
We remark that, for convenience, we have made a notational distinction
between the arbitrary state $[\psi ]$ and the eigenvectors of the operator $%
[A],$ which we denote by $[\xi ]$.

The eigenvalue equation (\ref{fi51}) is the counterpart of Eqn. (\ref{fo40})
but it contains extra detail which permits us to give a better
interpretation of the quantities that appear in it. The elements of the
eigenvector are the probability amplitudes $\chi (\lambda _k;B_j)$. In
general, these probability amplitudes describe measurements from the initial
state corresponding to the eigenvalue $\lambda _k$ to intermediate states
which are different from the ones appearing as the final states in the
differential eigenvalue equation. Thus, the matrix mechanics description
involves a complete set of intermediate states corresponding to an
intermediate observable $B$. This observable may be chosen in a variety of
ways; if it is chosen to equal the initial observable $A$, we obtain the
results corresponding to using the eigenrepresentation.

The fact that the elements of the eigenvector are probability amplitudes is
of course standard knowledge; however, we see here that these probability
amplitudes have a structure and are far from being mere constants. As a
result of this structure, we are able to obtain the eigenvectors of complex
matrices by inspection, as will be seen in later sections.

The present approach affords a unified treatment of measurements on quantum
systems. Whatever the measurement we wish to make, we begin by talking in
terms of the probability amplitudes for the various possible outcomes of the
measurement. These can then be used to make the transition to
matrix-mechanics. Since all results of measurement can be characterized by
the probabilities that are associated with obtaining the various values than
can result from the measurement, this approach should always be possible.
Therefore it is very general.

In this paper, this approach is used to treat spin. It permits us to derive
the spin vectors and operators from a more fundamental level of analysis
even though the spin cannot be expressed in terms of the position
coordinates.

\section{\bf SPECIALIZATION TO SPIN-1/2 SYSTEMS}

\subsection{\bf Spin-1/2 Systems}

Consider a spin-$1/2$ system whose spin ${\bf \sigma }$ is known to be up
with respect to the unit vector $\widehat{{\bf a}}$. A measurement of the
spin along a new unit vector $\widehat{{\bf c}}$ is made. At the same time,
a measurement of the quantity $R({\bf \sigma }\cdot \widehat{{\bf c}})\;$%
which is a function of the spin projection ${\bf \sigma \cdot }\widehat{{\bf %
c}}\;$is made. In order to describe the measurements, we introduce the
probability amplitudes for the measurements. These are the equivalent of the
wave function, except that they are discrete. We shall assume that all spins
are measured in units of $\hbar /2.$

We want to use the Land\'e expression Eqn. (\ref{nine}) to expand each of
the amplitudes. The initial and final states are spin states. The
intermediate state is also a spin state. This arises from the fact that spin
measurements along different quantization directions have the status of
different observables. The advantage of this circumstance is that though the
initial, intermediate states and final states formally have differing
statuses, they in practice are connected by probability amplitudes of the
same form.

Let the states be labelled by the magnetic quantum numbers. Since each such
number is with respect to a quantization direction, a superscript indicating
this direction is also needed. Thus, $\psi ((+\frac 12)^{(\widehat{{\bf a}}%
)};(+\frac 12)^{(\widehat{{\bf c}})})\;$is the probability amplitude for the
spin of the system being found upon measurement to be up with respect to $%
\widehat{{\bf c}}$, after the spin had previously been ascertained to be up
with respect to $\widehat{{\bf a}}$. There are three other probability
amplitudes; these are $\psi ((+\frac 12)^{(\widehat{{\bf a}})};(-\frac 12)^{(%
\widehat{{\bf c}})})$, $\psi ((-\frac 12)^{(\widehat{{\bf a}})};(+\frac
12)^{(\widehat{{\bf c}})})\;$and $\psi ((-\frac 12)^{(\widehat{{\bf a}}%
)};(-\frac 12)^{(\widehat{{\bf c}})})$, with obvious interpretations. We
note that orthogonality of the spin states corresponding to one quantization
direction means, for example, that 
\begin{equation}
\psi ((+\frac 12)^{(\widehat{{\bf a}})};(-\frac 12)^{(\widehat{{\bf a}}%
)})=\psi ((-\frac 12)^{(\widehat{{\bf a}})};(+\frac 12)^{(\widehat{{\bf a}}%
)})=0  \label{fi54a}
\end{equation}

According to Eqn. (\ref{nine}), we have the following expansions:

\begin{eqnarray}
\psi ((+\frac 12)^{(\widehat{{\bf a}})};(+\frac 12)^{(\widehat{{\bf c}})})
&=&\chi ((+\frac 12)^{(\widehat{{\bf a}})};(+\frac 12)^{(\widehat{{\bf b}}%
)})\phi ((+\frac 12)^{(\widehat{{\bf b}})};(+\frac 12)^{(\widehat{{\bf c}})})
\nonumber  \label{fi55} \\
&&+\chi ((+\frac 12)^{(\widehat{{\bf a}})};(-\frac 12)^{(\widehat{{\bf b}}%
)})\phi ((-\frac 12)^{(\widehat{{\bf b}})};(+\frac 12)^{(\widehat{{\bf c}})})
\label{fi55}
\end{eqnarray}

\begin{eqnarray}
\psi ((+\frac 12)^{(\widehat{{\bf a}})};(-\frac 12)^{(\widehat{{\bf c}})})
&=&\chi ((+\frac 12)^{(\widehat{{\bf a}})};(+\frac 12)^{(\widehat{{\bf b}}%
)})\phi ((+\frac 12)^{(\widehat{{\bf b}})};(-\frac 12)^{(\widehat{{\bf c}})})
\nonumber  \label{fi56} \\
&&\ +\chi ((+\frac 12)^{(\widehat{{\bf a}})};(-\frac 12)^{(\widehat{{\bf b}}%
)}))\phi ((-\frac 12)^{(\widehat{{\bf b}})};(-\frac 12)^{(\widehat{{\bf c}}%
)})  \label{fi56}
\end{eqnarray}
\begin{eqnarray}
\psi ((-\frac 12)^{(\widehat{{\bf a}})};(+\frac 12)^{(\widehat{{\bf c}})})
&=&\chi ((-\frac 12)^{(\widehat{{\bf a}})};(+\frac 12)^{(\widehat{{\bf b}}%
)})\phi ((+\frac 12)^{(\widehat{{\bf b}})};(+\frac 12)^{(\widehat{{\bf c}})})
\nonumber  \label{fi57} \\
&&\ +\chi ((-\frac 12)^{(\widehat{{\bf a}})};(-\frac 12)^{(\widehat{{\bf b}}%
)})\phi ((-\frac 12)^{(\widehat{{\bf b}})};(+\frac 12)^{(\widehat{{\bf c}})})
\label{fi57}
\end{eqnarray}
and 
\begin{eqnarray}
\psi ((-\frac 12)^{(\widehat{{\bf a}})};(-\frac 12)^{(\widehat{{\bf c}})})
&=&\chi ((-\frac 12)^{(\widehat{{\bf a}})};(+\frac 12)^{(\widehat{{\bf b}}%
)})\phi ((+\frac 12)^{(\widehat{{\bf b}})};(-\frac 12)^{(\widehat{{\bf c}})})
\nonumber \\
&&\ +\chi ((-\frac 12)^{(\widehat{{\bf a}})};(-\frac 12)^{(\widehat{{\bf b}}%
)})\phi ((-\frac 12)^{(\widehat{{\bf b}})};(-\frac 12)^{(\widehat{{\bf c}}%
)}).  \label{fi58}
\end{eqnarray}

Here $\widehat{{\bf b}}$ is another unit vector along which we can measure
the spin projection. The basis functions for the first and third expansions
are the probability amplitudes $\phi ((+\frac 12)^{(\widehat{{\bf b}}%
)};(+\frac 12)^{(\widehat{{\bf c}})})\;$and $\phi ((-\frac 12)^{(\widehat{%
{\bf b}})};(+\frac 12)^{(\widehat{{\bf c}})})$. For the second and fourth
expansions, they are $\phi ((+\frac 12)^{(\widehat{{\bf b}})};(-\frac 12)^{(%
\widehat{{\bf c}})})$ and $\phi ((-\frac 12)^{(\widehat{{\bf b}})};(-\frac
12)^{(\widehat{{\bf c}})})$.

According to Eqns. (\ref{fo46}) and (\ref{fi53}), the matrix representations
of these probability amplitudes $\psi $ are 
\begin{equation}
\lbrack \psi ((+\frac 12)^{(\widehat{{\bf a}})};(+\frac 12)^{(\widehat{{\bf c%
}})})]=[\psi ((+\frac 12)^{(\widehat{{\bf a}})};(-\frac 12)^{(\widehat{{\bf c%
}})})]=\left( 
\begin{array}{c}
\chi ((+\frac 12)^{(\widehat{{\bf a}})};(+\frac 12)^{(\widehat{{\bf b}})})
\\ 
\chi ((+\frac 12)^{(\widehat{{\bf a}})};(-\frac 12)^{(\widehat{{\bf b}})})
\end{array}
\right)  \label{fi59}
\end{equation}
and 
\begin{equation}
\lbrack \psi ((-\frac 12)^{(\widehat{{\bf a}})};(+\frac 12)^{(\widehat{{\bf c%
}})})]=[\psi ((-\frac 12)^{(\widehat{{\bf a}})};(-\frac 12)^{(\widehat{{\bf c%
}})})]=\left( 
\begin{array}{c}
\chi ((-\frac 12)^{(\widehat{{\bf a}})};(+\frac 12)^{(\widehat{{\bf b}})})
\\ 
\chi ((-\frac 12)^{(\widehat{{\bf a}})};(-\frac 12)^{(\widehat{{\bf b}})})
\end{array}
\right) .  \label{si60}
\end{equation}
We observe that provided the initial state is the same, the matrix state is
the same.

We now denote the quantum numbers as $m_1=+\frac 12$ and $m_2=-\frac 12$.
The general probability amplitude $\psi $ may therefore be denoted by $\psi
(m_i^{(\widehat{{\bf a}})};m_n^{(\widehat{{\bf c}})})$, where $(i,n=1,2).$
The same kind of notation will apply to the probability amplitudes $\chi $
and $\phi .$

Let the quantity $R({\bf \sigma }\cdot \widehat{{\bf c}})$ have the value $%
r_1$ when the spin projection is up (quantum number $m_1$) with respect to
the vector $\widehat{{\bf c}}$, and $r_{2\;}$when it is down (quantum number 
$m_2$). Thus, the possible values of $R$ are $r_n$ $(n=1,2)$. Suppose that
the initial state corresponds to the quantum number $m_i$ with respect to $%
\widehat{{\bf a}}$. As the probability amplitude for obtaining $r_n$ is $%
\psi (m_i^{(\widehat{{\bf a}})};m_n^{(\widehat{{\bf c}})})$, the expectation
value of $R$ is

\begin{equation}
\left\langle R\right\rangle =\sum_{n=1}^2\left| \psi (m_i^{(\widehat{{\bf a}}%
)};m_n^{(\widehat{{\bf c}})})\right| ^2r_n.  \label{si61}
\end{equation}

Since the expansions for $\psi ^{*}(m_i^{(\widehat{{\bf a}})};m_n^{(\widehat{%
{\bf c}})})$ and $\psi (m_i^{(\widehat{{\bf a}})};m_n^{(\widehat{{\bf c}})})$
are

\begin{equation}
\psi ^{*}(m_i^{(\widehat{{\bf a}})};m_n^{(\widehat{{\bf c}}%
)})=\sum_{j=1}^2\chi ^{*}(m_i^{(\widehat{{\bf a}})};m_j^{(\widehat{{\bf b}}%
)})\phi ^{*}(m_j^{(\widehat{{\bf b}})};m_n^{(\widehat{{\bf c}})})
\label{si62}
\end{equation}
and

\begin{equation}
\psi (m_i^{(\widehat{{\bf a}})};m_n^{(\widehat{{\bf c}})})=\sum_{j^{\prime
}=1}^2\chi (m_i^{(\widehat{{\bf a}})};m_{j^{\prime }}^{(\widehat{{\bf b}}%
)})\phi (m_{j^{\prime }}^{(\widehat{{\bf b}})};m_n^{(\widehat{{\bf c}})}),
\label{si63}
\end{equation}
it follows that

\begin{eqnarray}
\left\langle R\right\rangle &=&\sum_j\sum_{j^{\prime }}\chi ^{*}(m_i^{(%
\widehat{{\bf a}})};m_j^{(\widehat{{\bf b}})})R_{jj^{\prime }}\chi (m_i^{(%
\widehat{{\bf a}})};m_{j^{\prime }}^{(\widehat{{\bf b}})})  \nonumber
\label{si64} \\
&=&[\psi (m_i^{(\widehat{{\bf a}})};m_n^{(\widehat{{\bf c}})})]^{\dagger
}[R][\psi (m_i^{(\widehat{{\bf a}})};m_n^{(\widehat{{\bf c}})})]
\label{si64}
\end{eqnarray}
where 
\begin{equation}
\lbrack \psi (m_i^{(\widehat{{\bf a}})};m_n^{(\widehat{{\bf c}})})]=\left( 
\begin{array}{c}
\chi (m_i^{(\widehat{{\bf a}})};m_1^{(\widehat{{\bf b}})}) \\ 
\chi (m_i^{(\widehat{{\bf a}})};m_2^{(\widehat{{\bf b}})})
\end{array}
\right)  \label{si64a}
\end{equation}
and 
\begin{equation}
\left[ R\right] =\left( 
\begin{array}{cc}
R_{11} & R_{12} \\ 
R_{21} & R_{22}
\end{array}
\right) ,  \label{si65}
\end{equation}
with the elements of $[R]$ being given by

\begin{equation}
R_{jj^{\prime }}=\sum_{n=1}^2\phi ^{*}(m_j^{(\widehat{{\bf b}})};m_n^{(%
\widehat{{\bf c}})})\phi (m_{j^{\prime }}^{(\widehat{{\bf b}})};m_n^{(%
\widehat{{\bf c}})})r_n  \label{si65a}
\end{equation}

Written out explicitly, the elements of $\;\left[ R\right] \;$are

\begin{equation}
R_{11}=\left| \phi ((+\frac 12)^{(\widehat{{\bf b}})};(+\frac 12)^{(\widehat{%
{\bf c}})})\right| ^2r_1+\left| \phi ((+\frac 12)^{(\widehat{{\bf b}}%
)};(-\frac 12)^{(\widehat{{\bf c}})})\right| ^2r_2,  \label{si66}
\end{equation}

\begin{eqnarray}
R_{12} &=&\phi ^{*}((+\frac 12)^{(\widehat{{\bf b}})};(+\frac 12)^{(\widehat{%
{\bf c}})})\phi ((-\frac 12)^{(\widehat{{\bf b}})};(+\frac 12)^{(\widehat{%
{\bf c}})})r_1  \nonumber  \label{si67} \\
&&+\phi ^{*}((+\frac 12)^{(\widehat{{\bf b}})};(-\frac 12)^{(\widehat{{\bf c}%
})})\phi ((-\frac 12)^{(\widehat{{\bf b}})};(-\frac 12)^{(\widehat{{\bf c}}%
)})r_2,  \label{si67}
\end{eqnarray}

\begin{eqnarray}
R_{21} &=&\phi ^{*}((-\frac 12)^{(\widehat{{\bf b}})};(+\frac 12)^{(\widehat{%
{\bf c}})})\phi ((+\frac 12)^{(\widehat{{\bf b}})};(+\frac 12)^{(\widehat{%
{\bf c}})})r_1  \nonumber \\
&&+\phi ^{*}((-\frac 12)^{(\widehat{{\bf b}})};(-\frac 12)^{(\widehat{{\bf c}%
})})\phi ((+\frac 12)^{(\widehat{{\bf b}})};(-\frac 12)^{(\widehat{{\bf c}}%
)})r_2  \label{si68}
\end{eqnarray}
and

\begin{equation}
R_{22}=\left| \phi ((-\frac 12)^{(\widehat{{\bf b}})};(+\frac 12)^{(\widehat{%
{\bf c}})})\right| ^2r_1+\left| \phi ((-\frac 12)^{(\widehat{{\bf b}}%
)};(-\frac 12)^{(\widehat{{\bf c}})})\right| ^2r_2.  \label{si69}
\end{equation}

When $R={\bf \sigma }\cdot \widehat{{\bf c}}$, then it is the component of
the spin in the direction $\widehat{{\bf c}}$. Hence, $\left[ R\right] $ is
the matrix form of the operator for the component of the spin along the axis
defined by $\widehat{{\bf c}}$. When $R=\sigma ^2$, then $[R]$ is the matrix
form of the square of the total spin.

\subsection{\bf Summary of the Various Possible Choices for the Reference
Vectors}

The results given above are the most general possible. We now consider
several different cases that arise when we choose the vectors $\widehat{{\bf %
b}}$ and $\widehat{{\bf c}}$ in specific ways.

Case (a): $\widehat{{\bf b}}\neq \widehat{{\bf a}}$ and $\widehat{{\bf c}}%
\neq \widehat{{\bf a}}$.

This is the most general case. In this case, the matrix representations are 
\begin{equation}
\lbrack \psi ((+\frac 12)^{(\widehat{{\bf a}})};(+\frac 12)^{(\widehat{{\bf c%
}})})]=[\psi ((+\frac 12)^{(\widehat{{\bf a}})};(-\frac 12)^{(\widehat{{\bf c%
}})})]=\left( 
\begin{array}{c}
\chi ((+\frac 12)^{(\widehat{{\bf a}})};(+\frac 12)^{(\widehat{{\bf b}})})
\\ 
\chi ((+\frac 12)^{(\widehat{{\bf a}})};(-\frac 12)^{(\widehat{{\bf b}})})
\end{array}
\right)  \label{si67z}
\end{equation}
and 
\begin{equation}
\lbrack \psi ((-\frac 12)^{(\widehat{{\bf a}})};(+\frac 12)^{(\widehat{{\bf c%
}})})]=[\psi ((-\frac 12)^{(\widehat{{\bf a}})};(-\frac 12)^{(\widehat{{\bf c%
}})})]=\left( 
\begin{array}{c}
\chi ((-\frac 12)^{(\widehat{{\bf a}})};(+\frac 12)^{(\widehat{{\bf b}})})
\\ 
\chi ((-\frac 12)^{(\widehat{{\bf a}})};(-\frac 12)^{(\widehat{{\bf b}})})
\end{array}
\right)  \label{si68z}
\end{equation}
while the operator $\left[ R\right] $ has its elements given by Eqns. (\ref
{si66}) - (\ref{si69}).

{\bf Case (b)}: $\widehat{{\bf b}}=\widehat{{\bf a}}.$

For this case, the matrix representations are 
\begin{equation}
\lbrack \psi ((+\frac 12)^{(\widehat{{\bf a}})};(+\frac 12)^{(\widehat{{\bf c%
}})})]=[\psi ((+\frac 12)^{(\widehat{{\bf a}})};(-\frac 12)^{(\widehat{{\bf c%
}})})]=\left( 
\begin{array}{c}
1 \\ 
0
\end{array}
\right)  \label{si69z}
\end{equation}
and

\begin{equation}
\lbrack \psi ((-\frac 12)^{(\widehat{{\bf a}})};(+\frac 12)^{(\widehat{{\bf c%
}})})]=[\psi ((-\frac 12)^{(\widehat{{\bf a}})};(-\frac 12)^{(\widehat{{\bf c%
}})})]=\left( 
\begin{array}{c}
0 \\ 
1
\end{array}
\right)  \label{se70}
\end{equation}
while the elements of $\left[ R\right] $ are Eqns. (\ref{si66}) - (\ref{si69}%
) with $\widehat{{\bf b}}$ replaced by $\widehat{{\bf a}}$.

{\bf Case (c)}: $\widehat{{\bf b}}=\widehat{{\bf c}}.$

The matrix representations are 
\begin{equation}
\lbrack \psi ((+\frac 12)^{(\widehat{{\bf a}})};(+\frac 12)^{(\widehat{{\bf c%
}})})]=[\psi ((+\frac 12)^{(\widehat{{\bf a}})};(-\frac 12)^{(\widehat{{\bf c%
}})})]=\left( 
\begin{array}{c}
\chi ((+\frac 12)^{(\widehat{{\bf a}})};(+\frac 12)^{(\widehat{{\bf c}})})
\\ 
\chi ((+\frac 12)^{(\widehat{{\bf a}})};(-\frac 12)^{(\widehat{{\bf c}})})
\end{array}
\right)  \label{se75}
\end{equation}
and

\begin{equation}
\lbrack \psi ((-\frac 12)^{(\widehat{{\bf a}})};(+\frac 12)^{(\widehat{{\bf c%
}})})]=[\psi ((-\frac 12)^{(\widehat{{\bf a}})};(-\frac 12)^{(\widehat{{\bf c%
}})})]=\left( 
\begin{array}{c}
\chi ((-\frac 12)^{(\widehat{{\bf a}})};(+\frac 12)^{(\widehat{{\bf c}})})
\\ 
\chi ((-\frac 12)^{(\widehat{{\bf a}})};(-\frac 12)^{(\widehat{{\bf c}})})
\end{array}
\right)  \label{se76}
\end{equation}
while the operator is

\begin{equation}
\left[ R\right] =\left( 
\begin{array}{cc}
r_1 & 0 \\ 
0 & r_2
\end{array}
\right) .  \label{se77}
\end{equation}

{\bf Case (d)}: $\widehat{{\bf c}}=\widehat{{\bf a}}$.

The matrix representations are

\begin{equation}
\lbrack \psi ((+\frac 12)^{(\widehat{{\bf a}})};(+\frac 12)^{(\widehat{{\bf a%
}})})]=[\psi ((+\frac 12)^{(\widehat{{\bf a}})};(-\frac 12)^{(\widehat{{\bf a%
}})})]=\left( 
\begin{array}{c}
\chi ((+\frac 12)^{(\widehat{{\bf a}})};(+\frac 12)^{(\widehat{{\bf b}})})
\\ 
\chi ((+\frac 12)^{(\widehat{{\bf a}})};(-\frac 12)^{(\widehat{{\bf b}})})
\end{array}
\right)  \label{se78}
\end{equation}
and 
\begin{equation}
\lbrack \psi ((-\frac 12)^{(\widehat{{\bf a}})};(+\frac 12)^{(\widehat{{\bf a%
}})})]=[\psi ((-\frac 12)^{(\widehat{{\bf a}})};(-\frac 12)^{(\widehat{{\bf a%
}})})]=\left( 
\begin{array}{c}
\chi ((-\frac 12)^{(\widehat{{\bf a}})};(+\frac 12)^{(\widehat{{\bf b}})})
\\ 
\chi ((-\frac 12)^{(\widehat{{\bf a}})};(-\frac 12)^{(\widehat{{\bf b}})})
\end{array}
\right)  \label{se79}
\end{equation}
while the elements of $\left[ R\right] $ are given by Eqns. (\ref{si66}) - (%
\ref{si69}) with $\widehat{{\bf c}}$ replaced by $\widehat{{\bf a}}$. We
observe that despite the generalized form of the matrix representations, the
probability amplitudes themselves are 
\begin{equation}
\psi ((+\frac 12)^{(\widehat{{\bf a}})};(+\frac 12)^{(\widehat{{\bf a}}%
)})=\psi ((-\frac 12)^{(\widehat{{\bf a}})};(-\frac 12)^{(\widehat{{\bf a}}%
)}=1  \label{se79x}
\end{equation}
and 
\begin{equation}
\psi ((+\frac 12)^{(\widehat{{\bf a}})};(-\frac 12)^{(\widehat{{\bf a}}%
)})=\psi ((-\frac 12)^{(\widehat{{\bf a}})};(+\frac 12)^{(\widehat{{\bf a}}%
)})=0.  \label{se79z}
\end{equation}
$.$

{\bf Case (e}): $\widehat{{\bf b}}=\widehat{{\bf a}}$ and $\widehat{{\bf c}}=%
\widehat{{\bf a}}$.

The matrix representations are

\begin{equation}
\lbrack \psi ((+\frac 12)^{(\widehat{{\bf a}})};(+\frac 12)^{(\widehat{{\bf a%
}})})]=[\psi ((+\frac 12)^{(\widehat{{\bf a}})};(-\frac 12)^{(\widehat{{\bf a%
}})})]=\left( 
\begin{array}{c}
1 \\ 
0
\end{array}
\right)  \label{ei80}
\end{equation}
and

\begin{equation}
\lbrack \psi ((-\frac 12)^{(\widehat{{\bf a}})};(+\frac 12)^{(\widehat{{\bf a%
}})})]=[\psi ((-\frac 12)^{(\widehat{{\bf a}})};(-\frac 12)^{(\widehat{{\bf a%
}})})]=\left( 
\begin{array}{c}
0 \\ 
1
\end{array}
\right)  \label{ei81}
\end{equation}
while the operator is

\begin{equation}
\left[ R\right] =\left( 
\begin{array}{cc}
r_1 & 0 \\ 
0 & r_2
\end{array}
\right) .  \label{ei82}
\end{equation}

We observe that Cases (a)-(c) are just different ways of treating exactly
the same measurements and computing the same expectation value.The initial
projection axis and the same final projection axis are all the same; the
situation corresponds to the general case. On the other hand, Cases (d) and
(e), which are also equivalent, both refer to a special case. Since $%
\widehat{{\bf a}}=\widehat{{\bf c}}$, both describe a repetition of the spin
projection measurement along a given axis.

\subsection{\bf Derivation of the Pauli Spin Matrices and Vectors }

The expressions for $[R]$ allow us to deduce the matrix forms of various
operators. Thus if $R$ is the spin projection itself, then $r_1=1\;$and $%
r_2=-1$, so that from Case (e), we are able to deduce the operator for the
spin projection along the axis of quantization as

\begin{equation}
\left[ \sigma _z\right] =\left( 
\begin{array}{cc}
1 & 0 \\ 
0 & -1
\end{array}
\right) .  \label{ei83}
\end{equation}
Eqn. (\ref{ei83}) is of course the Pauli matrix for the $z$ component of the
spin.

Even though the $z$ axis has not been explicitly introduced, we are using $z$
as a subscript for a good reason. The generalized form of this operator, to
be derived below, will be defined with respect to an arbitrary unit vector,
say $\widehat{{\bf c}}$. It will be seen that in an appropriate limit, this
generalized operator will reduce to the Pauli matrix Eqn. (\ref{ei83}). This
circumstance justifies our use of the subscript $z$.

Now, since the eigenvalue of $\sigma ^2$ is 3, we have $r_1=r_2=3.$
Therefore the operator for the square of the spin is given by Case (e) as

\begin{equation}
\left[ \sigma ^2\right] =\left( 
\begin{array}{cc}
3 & 0 \\ 
0 & 3
\end{array}
\right)  \label{ei83z}
\end{equation}

We emphasize that these operators apply when $\widehat{{\bf b}}=\widehat{%
{\bf c}}$ and when $\widehat{{\bf a}}=\widehat{{\bf b}}=\widehat{{\bf c}}.$

Once we have the $z$ component and its eigenvectors, we invoke standard
theory to obtain the $x$ and $y$ components. Thus, we derive the matrix
forms of the operators for the $x$ and $y$ components of spin by first
obtaining the ladder operators $\left[ \sigma _{+}\right] \;$and $\left[
\sigma _{-}\right] .$ These are deduced from their actions on the
eigenvectors of $[\sigma _z]$. The eigenvectors of $[\sigma _z]$ are $[\xi
_{+}]=\left( 
\begin{array}{c}
1 \\ 
0
\end{array}
\right) $ corresponding to the eigenvalue $+1$, and $[\xi _{-}]=\left( 
\begin{array}{c}
0 \\ 
1
\end{array}
\right) $corresponding to the eigenvalue $-1$.

In terms of the ladder operators, the operators for the $x$ and $y$
components of the spin are

\begin{equation}
\lbrack \sigma _x]=\frac 12([\sigma _{+}]+[\sigma _{-}])  \label{ei84}
\end{equation}
and

\begin{equation}
\lbrack \sigma _y]=-\frac i2([\sigma _{+}]-[\sigma _{-}]).  \label{ei85}
\end{equation}

The effects of $[\sigma _{+}]$ on the eigenvectors of $[\sigma _z]$ are

\begin{equation}
\lbrack \sigma _{+}][\xi _{+}]=0  \label{ei86}
\end{equation}
and 
\begin{equation}
\left[ \sigma _{+}\right] [\xi _{-}]=2[\xi _{+}].  \label{ei87}
\end{equation}
Using these properties, we find that

\begin{equation}
\left[ \sigma _{+}\right] =2\left( 
\begin{array}{cc}
0 & 1 \\ 
0 & 0
\end{array}
\right) .  \label{ei88}
\end{equation}
Similarly, using

\begin{equation}
\left[ \sigma _{-}\right] [\xi _{-}]=0  \label{ei89}
\end{equation}
and 
\begin{equation}
\left[ \sigma _{-}\right] [\xi _{-}]=2[\xi _{+}]  \label{ni90}
\end{equation}
we find that

\begin{equation}
\left[ \sigma _{-}\right] =2\left( 
\begin{array}{cc}
0 & 0 \\ 
1 & 0
\end{array}
\right) .  \label{ni91}
\end{equation}
Hence

\begin{equation}
\left[ \sigma _x\right] =\left( 
\begin{array}{cc}
0 & 1 \\ 
1 & 0
\end{array}
\right)  \label{ni92}
\end{equation}
and 
\begin{equation}
\lbrack \sigma _y]=\left( 
\begin{array}{cc}
0 & -i \\ 
i & 0
\end{array}
\right) .  \label{ni93}
\end{equation}

These are the Pauli spin matrices, which we have here derived from first
principles. We see that the Pauli spin vectors and matrices arise as a
special case of spin measurements. They result if we make a second
measurement of spin without changing the reference direction, and while
using the initial/final direction to define the intermediate states which
appear in the expansion of the relevant probability amplitudes by Eqn. (\ref
{nine}).

We need to obtain the most general possible vectors and operators for the
description of spin. In order to obtain these quantities, we need to obtain
explicit expressions for the probability amplitudes $\chi $ and $\phi .$

\subsection{\bf Explicit Formulas for the Probability Amplitudes}

In the last section, we have obtained the matrix operator for the spin, for
the special case corresponding to the Pauli vectors and matrices; this
operator is

\begin{equation}
\lbrack {\bf \sigma ]}=\widehat{{\bf i}}\left( 
\begin{array}{cc}
0 & 1 \\ 
1 & 0
\end{array}
\right) +\widehat{{\bf j}}\left( 
\begin{array}{cc}
0 & -i \\ 
i & 0
\end{array}
\right) +\widehat{{\bf k}}\left( 
\begin{array}{cc}
1 & 0 \\ 
0 & -1
\end{array}
\right) .  \label{ni94}
\end{equation}

Consider the operator obtained from ${\bf \sigma }$ by taking its dot
product with a unit vector $\widehat{{\bf a}}$. A measurement of spin along
the new direction defined by the unit vector ${\bf a}$ gives the values $\pm
1.$ Let the polar angles of $\widehat{{\bf a}}$ be $(\theta ,\varphi )$, so
that $\widehat{{\bf a}}=(\sin \theta \cos \varphi ,\sin \theta \sin \varphi
,\cos \theta ).$ The component of the spin along $\widehat{{\bf a}}$ is $%
{\bf \sigma }\cdot \widehat{{\bf a}}$. In view of Eqn. (\ref{ni94}) and of
the components of $\widehat{{\bf a}}$, the matrix form of this operator is

\begin{equation}
\lbrack {\bf \sigma }\cdot \widehat{{\bf a}}]=\left( 
\begin{array}{cc}
\cos \theta & \sin \theta e^{-i\varphi } \\ 
\sin \theta e^{i\varphi } & -\cos \theta
\end{array}
\right) .  \label{ni95}
\end{equation}
The eigenvalues of this operator are $+1$ with eigenvector

\begin{equation}
\lbrack \xi _{+}]=\left( 
\begin{array}{c}
\cos \theta /2 \\ 
e^{i\varphi }\sin \theta /2
\end{array}
\right)  \label{ni96}
\end{equation}
and $-1$ with eigenvector

\begin{equation}
\lbrack \xi _{-}]=\left( 
\begin{array}{c}
\sin \theta /2 \\ 
-e^{i\varphi }\cos \theta /2
\end{array}
\right) .  \label{ni97}
\end{equation}

From the matrix eigenvalue equation (\ref{fi51}), we deduce that the
elements of the eigenvectors are of the form $\chi (\lambda _k;B_j),$ where $%
\lambda _k$ is an eigenvalue of the matrix, and corresponds to the state
that precedes measurement, while $B_j$ is an eigenvalue corresponding to
another state. Thus, these probability amplitudes have the structure of the $%
\chi (\lambda _k;B_j).$

In general $B$ is not the observable being measured and is also different
from the observable corresponding to the initial state. But for the case of
spin measurements, provided that $B$ represents spin projection measurements
along a new direction, it has the status of another observable. We therefore
conclude that in the vectors $[\xi _1]$ and $[\xi _2],$ Eqns. (\ref{ni96})
and (\ref{ni97}) respectively, the elements are probability amplitudes which
refer to measurements from an initial state in which the spin is up or down
with respect to $\widehat{{\bf a}}$ . In $[\xi _1]$, the elements are
probability amplitudes corresponding to an initial state characterized by
the spin projection being up with respect to $\widehat{{\bf a}};$ the final
states correspond to the eigenvalues $+1$ and $-1$ with respect to the unit
vector $\widehat{{\bf b}}$. In $[\xi _2]$, the elements are probability
amplitudes corresponding to the initial state characterized by the spin
projection being down with respect to $\widehat{{\bf a}}$; the final states
correspond to the eigenvalues $+1$ and $-1$ with respect to the unit vector $%
\widehat{{\bf b}}$. Hence we deduce from Eqn. (\ref{ni96}) that

\begin{equation}
\chi ((+\frac 12)^{(\widehat{{\bf a}})};(+\frac 12)^{(\widehat{{\bf b}}%
)})=\cos \theta /2  \label{ni98}
\end{equation}
and

\begin{equation}
\chi ((+\frac 12)^{(\widehat{{\bf a}})};(-\frac 12)^{(\widehat{{\bf b}}%
)})=e^{i\varphi }\sin \theta /2.  \label{ni99}
\end{equation}

Similarly, we deduce from Eqn. \ref{ni97}$\;$that 
\begin{equation}
\chi ((-\frac 12)^{(\widehat{{\bf a}})};(+\frac 12)^{(\widehat{{\bf b}}%
)})=\sin \theta /2.  \label{hu100}
\end{equation}
and 
\begin{equation}
\chi ((-\frac 12)^{(\widehat{{\bf a}})};(-\frac 12)^{(\widehat{{\bf b}}%
)})=-e^{i\varphi }\cos \theta /2.  \label{hu101}
\end{equation}

For the time being, we do not need to know what $\widehat{{\bf b}}$ is,
though we shall deduce it later. As $\widehat{{\bf b}}$ may be a special
vector, we cannot assume that the probability amplitudes Eqns. (\ref{ni98})
- (\ref{hu101}) are general. But armed with these special amplitudes, we are
able to obtain general expressions for the probability amplitudes using Eqn.
(\ref{nine}).

At this point, we note a simplification which arises from the fact that in
all the probability amplitudes under consideration, the initial and final
states both correspond to spin projections. This means that in reality, the
probability amplitudes $\psi $, $\chi $ and $\phi $ have identical forms.

Let $\widehat{{\bf c}}$ be a new unit vector defined by the polar angles $%
(\theta ^{\prime },\varphi ^{\prime }).$ Then if we make measurements of
spin from $\widehat{{\bf c}}$ to the unknown quantization direction $%
\widehat{{\bf b}}$, the probability amplitudes are

\begin{equation}
\chi ((+\frac 12)^{(\widehat{{\bf c}})};(+\frac 12)^{(\widehat{{\bf b}}%
)})=\cos \theta ^{\prime }/2  \label{hu102}
\end{equation}

\begin{equation}
\chi ((+\frac 12)^{(\widehat{{\bf c}})};(-\frac 12)^{(\widehat{{\bf b}}%
)})=e^{i\varphi ^{\prime }}\sin \theta ^{\prime }/2  \label{hu103}
\end{equation}
\begin{equation}
\chi ((-\frac 12)^{(\widehat{{\bf c}})};(+\frac 12)^{(\widehat{{\bf b}}%
)})=\sin \theta ^{\prime }/2  \label{hu104}
\end{equation}
and 
\begin{equation}
\chi ((-\frac 12)^{(\widehat{{\bf c}})};(-\frac 12)^{(\widehat{{\bf b}}%
)})=-e^{i\varphi ^{\prime }}\cos \theta ^{\prime }/2.  \label{hu105}
\end{equation}

We can now use Eqn. (\ref{nine}) to eliminate the unknown vector $\widehat{%
{\bf b}}$, and thereby find the general probability amplitudes $\chi $.
Using the Hermiticity condition Eqn. (\ref{eit}), we find that

\begin{eqnarray}
\chi ((+\frac 12)^{(\widehat{{\bf a}})};(+\frac 12)^{(\widehat{{\bf c}})})
&=&\chi ((+\frac 12)^{(\widehat{{\bf a}})};(+\frac 12)^{(\widehat{{\bf b}}%
)})\chi ((+\frac 12)^{(\widehat{{\bf b}})};(+\frac 12)^{(\widehat{{\bf c}})})
\nonumber  \label{hu106} \\
&&+\chi ((+\frac 12)^{(\widehat{{\bf a}})};(-\frac 12)^{(\widehat{{\bf b}}%
)})\chi ((-\frac 12)^{(\widehat{{\bf b}})};(+\frac 12)^{(\widehat{{\bf c}})})
\nonumber \\
&=&\cos \theta /2\cos \theta ^{\prime }/2+e^{i(\varphi -\varphi ^{\prime
})}\sin \theta /2\sin \theta ^{\prime }/2.  \label{hu106}
\end{eqnarray}
Similarly, we find that

\begin{equation}
\chi ((+\frac 12)^{(\widehat{{\bf a}})};(-\frac 12)^{(\widehat{{\bf c}}%
)})=\cos \theta /2\sin \theta ^{\prime }/2-e^{i(\varphi -\varphi ^{\prime
})}\sin \theta /2\cos \theta ^{\prime }/2  \label{hu107}
\end{equation}

\begin{equation}
\chi ((-\frac 12)^{(\widehat{{\bf a}})};(+\frac 12)^{(\widehat{{\bf c}}%
)})=\sin \theta /2\cos \theta ^{\prime }/2-e^{i(\varphi -\varphi ^{\prime
})}\cos \theta /2\sin \theta ^{\prime }/2  \label{hu108}
\end{equation}
and

\begin{equation}
\chi ((-\frac 12)^{(\widehat{{\bf a}})};(-\frac 12)^{(\widehat{{\bf c}}%
)})=\sin \theta /2\sin \theta ^{\prime }/2+e^{i(\varphi -\varphi ^{\prime
})}\cos \theta /2\cos \theta ^{\prime }/2.  \label{hu109}
\end{equation}
The expressions Eqns. (\ref{hu106}) - (\ref{hu109}) are the general
expressions for the probability amplitudes which we seek.

\subsection{\bf Verification of the General Expressions}

We need to check that the expressions for the probability amplitudes we have
obtained are correct. If they are, then they should satisfy the Land\'e
expansion Eqn. (\ref{nine}). Let $\widehat{{\bf e}}$ be a new unit vector
defined by the polar angles $(\theta ^{\prime \prime },\varphi ^{\prime
\prime }).$ Then according to Eqns. (\ref{hu106}) - (\ref{hu109}),

\begin{equation}
\chi ((+\frac 12)^{(\widehat{{\bf a}})};(+\frac 12)^{(\widehat{{\bf e}}%
)})=\cos \theta /2\cos \theta ^{\prime \prime }/2+e^{i(\varphi -\varphi
^{\prime \prime })}\sin \theta /2\sin \theta ^{\prime \prime }/2
\label{hu117a}
\end{equation}

\begin{equation}
\chi ((+\frac 12)^{(\widehat{{\bf a}})};(-\frac 12)^{(\widehat{{\bf e}}%
)})=\cos \theta /2\sin \theta ^{\prime \prime }/2-e^{i(\varphi -\varphi
^{\prime \prime })}\sin \theta /2\cos \theta ^{\prime \prime }/2
\label{hu117b}
\end{equation}

\begin{equation}
\chi ((-\frac 12)^{(\widehat{{\bf a}})};(+\frac 12)^{(\widehat{{\bf e}}%
)})=\sin \theta /2\cos \theta ^{\prime \prime }/2-e^{i(\varphi -\varphi
^{\prime \prime })}\cos \theta /2\sin \theta ^{\prime \prime }/2
\label{hu117c}
\end{equation}
and 
\begin{equation}
\chi ((-\frac 12)^{(\widehat{{\bf a}})};(-\frac 12)^{(\widehat{{\bf e}}%
)})=\sin \theta /2\sin \theta "/2+e^{i(\varphi -\varphi ^{^{\prime \prime
}})}\cos \theta /2\cos \theta ^{^{\prime \prime }}/2.  \label{hu117d}
\end{equation}

Also, 
\begin{equation}
\chi ((+\frac 12)^{(\widehat{{\bf e}})};(+\frac 12)^{(\widehat{{\bf c}}%
)})=\cos \theta ^{\prime \prime }/2\cos \theta ^{\prime }/2+e^{i(\varphi
^{\prime \prime }-\varphi ^{\prime })}\sin \theta ^{\prime \prime }/2\sin
\theta ^{\prime }/2  \label{hu117e}
\end{equation}

\begin{equation}
\chi ((+\frac 12)^{(\widehat{{\bf e}})};(-\frac 12)^{(\widehat{{\bf c}}%
)})=\cos \theta ^{\prime \prime }/2\sin \theta ^{\prime }/2-e^{i(\varphi
^{\prime \prime }-\varphi ^{\prime })}\sin ^{\prime \prime }\theta /2\cos
\theta ^{\prime }/2  \label{hu117f}
\end{equation}

\begin{equation}
\chi ((-\frac 12)^{(\widehat{{\bf e}})};(+\frac 12)^{(\widehat{{\bf c}}%
)})=\sin \theta ^{\prime \prime }/2\cos \theta ^{\prime }/2-e^{i(\varphi
^{\prime \prime }-\varphi ^{\prime })}\cos \theta ^{^{\prime \prime }}/2\sin
\theta ^{\prime }/2  \label{hu117g}
\end{equation}

\begin{equation}
\chi ((-\frac 12)^{(\widehat{{\bf e}})};(-\frac 12)^{(\widehat{{\bf c}}%
)})=\sin \theta ^{^{\prime \prime }}/2\sin \theta ^{\prime }/2+e^{i(\varphi
^{\prime \prime }-\varphi ^{\prime })}\cos \theta ^{\prime \prime }/2\cos
\theta ^{\prime }/2.  \label{hu117h}
\end{equation}

Now, 
\begin{eqnarray}
\chi ((+\frac 12)^{(\widehat{{\bf a}})};(+\frac 12)^{(\widehat{{\bf c}})})
&=&\chi ((+\frac 12)^{(\widehat{{\bf a}})};(+\frac 12)^{(\widehat{{\bf e}}%
)})\chi ((+\frac 12)^{(\widehat{{\bf e}})};(+\frac 12)^{(\widehat{{\bf c}})})
\nonumber \\
&&+\chi ((+\frac 12)^{(\widehat{{\bf a}})};(-\frac 12)^{(\widehat{{\bf e}}%
)})\chi ((-\frac 12)^{(\widehat{{\bf e}})};(+\frac 12)^{(\widehat{{\bf c}})})
\nonumber \\
&=&\cos \theta /2\cos \theta ^{\prime }/2+e^{i(\varphi -\varphi ^{\prime
})}\sin \theta /2\sin \theta ^{\prime }/2.  \label{hu117j}
\end{eqnarray}

When we substitute for the various probability amplitudes from the list
Eqns. (\ref{hu117a}) - (\ref{hu117h}), we recover Eqn. (\ref{hu106}). The
same holds for the other probability amplitudes Eqns. (\ref{hu107}) - (\ref
{hu109}). This confirms the consistency of the expressions for the
probability amplitudes and the validity of the Land\'e expansion Eqn. (\ref
{nine}).

\subsection{\bf Expressions for the Probabilities}

The probabilities corresponding to the probability amplitudes Eqns. (\ref
{hu106}) - (\ref{hu109}) are 
\begin{eqnarray}
P((+\frac 12)^{(\widehat{{\bf a}})};(+\frac 12)^{(\widehat{{\bf c}})})
&=&\left| \chi ((+\frac 12)^{(\widehat{{\bf a}})};(+\frac 12)^{(\widehat{%
{\bf c}})})\right| ^2  \nonumber  \label{hu110} \\
&=&\cos ^2(\theta /2-\theta ^{\prime }/2)-\sin \theta \sin \theta ^{\prime
}\sin ^2(\varphi /2-\varphi ^{\prime }/2)  \label{hu110}
\end{eqnarray}
\begin{equation}
P((+\frac 12)^{(\widehat{{\bf a}})};(-\frac 12)^{(\widehat{{\bf c}})})=\sin
^2(\theta /2-\theta ^{\prime }/2)+\sin \theta \sin \theta ^{\prime }\sin
^2(\varphi /2-\varphi ^{\prime }/2)  \label{hu111}
\end{equation}
\begin{equation}
P((-\frac 12)^{(\widehat{{\bf a}})};(+\frac 12)^{(\widehat{{\bf c}})})=\sin
^2(\theta /2-\theta ^{\prime }/2)+\sin \theta \sin \theta ^{\prime }\sin
^2(\varphi /2-\varphi ^{\prime }/2)  \label{hu112}
\end{equation}
and 
\begin{equation}
P((-\frac 12)^{(\widehat{{\bf a}})};(-\frac 12)^{(\widehat{{\bf c}})})=\cos
^2(\theta /2-\theta ^{\prime }/2)-\sin \theta \sin \theta ^{\prime }\sin
^2(\varphi /2-\varphi ^{\prime }/2).  \label{hu113}
\end{equation}

The cosine of the angle $\Theta $ between the vectors $\widehat{{\bf a}}$
and $\widehat{{\bf c}}$ is 
\begin{equation}
\cos \Theta =\widehat{{\bf a}}\cdot \widehat{{\bf c}}=\cos (\theta -\theta
^{\prime })-2\sin \theta \sin \theta ^{\prime }\sin ^2(\varphi -\varphi
^{\prime })/2.  \label{hu113a}
\end{equation}
Hence, 
\begin{equation}
\cos ^2\Theta /2=\cos ^2(\theta /2-\theta ^{\prime }/2)-\sin \theta \sin
\theta ^{\prime }\sin ^2(\varphi /2-\varphi ^{\prime }/2)  \label{hu113b}
\end{equation}
We see therefore that

\begin{equation}
P((+\frac 12)^{(\widehat{{\bf a}})};(+\frac 12)^{(\widehat{{\bf c}}%
)}))=P((-\frac 12)^{(\widehat{{\bf a}})};(-\frac 12)^{(\widehat{{\bf c}}%
)})=\cos ^2\Theta /2.  \label{hu113c}
\end{equation}
In the same fashion, we deduce that

\begin{equation}
P((+\frac 12)^{(\widehat{{\bf a}})};(-\frac 12)^{(\widehat{{\bf c}}%
)})=P((-\frac 12)^{(\widehat{{\bf a}})},(+\frac 12)^{(\widehat{{\bf c}}%
)})=\sin ^2\Theta /2.  \label{hu113d}
\end{equation}
As we expect, 
\begin{equation}
P((\pm \frac 12)^{(\widehat{{\bf a}})},(+\frac 12)^{(\widehat{{\bf c}}%
)})+P((\pm \frac 12)^{(\widehat{{\bf a}})};(-\frac 12)^{(\widehat{{\bf c}}%
)})=1.  \label{hu113s}
\end{equation}
Thus, the probabilities are normalized to unity as we expect.

We are now in a position to deduce the vector $\widehat{{\bf b}}$. It is
straightforward to see that $\widehat{{\bf b}}$ is the vector with angles $%
\theta ^{\prime }=0$ and $\varphi ^{\prime }=\pi .$ We expect therefore that
if we make these substitutions in the generalized probability amplitudes
Eqns. (\ref{hu106}) - (\ref{hu109}), we should recover the specialized
amplitudes Eqns. (\ref{ni98}) - (\ref{hu101}). This is precisely the case.

\subsection{\bf Explicit Formulas For the Operators}

We can now obtain explicit general formulas for the various operators. We
want to obtain the explicit expressions Eqns. (\ref{si66}) - (\ref{si69})
which are the elements of the general operator $\left[ R\right] $. Since
these are given in terms of vectors $\widehat{{\bf b}}$ and $\widehat{{\bf c}%
}$, we have to redefine the angles so that they refer to these vectors. We
define the unit vector $\widehat{{\bf b}}$ by the angles $(\theta ,\varphi )$%
, while the unit vector $\widehat{{\bf c}}$ is defined by the angles $%
(\theta ^{\prime },\varphi ^{\prime })$. It is important to realize that the
vector $\widehat{{\bf b}}$ in this subsection is different from the vector $%
\widehat{{\bf b}}$ in Subsections $4.4$ and $4.6$. The vector in Subsections 
$4.4$ and $4.6$ is a particular instance of the general vector in this
section.

Using the fact that the $\phi $'s and $\chi $'s have identical forms, we
find that

\begin{equation}
\phi ((+\frac 12)^{(\widehat{{\bf b}})};(+\frac 12)^{(\widehat{{\bf c}}%
)})=\cos \theta /2\cos \theta ^{\prime }/2+e^{i(\varphi -\varphi ^{\prime
})}\sin \theta /2\sin \theta ^{\prime }/2,  \label{hu118}
\end{equation}

\begin{equation}
\phi ((+\frac 12)^{(\widehat{{\bf b}})};(-\frac 12)^{(\widehat{{\bf c}}%
)})=\cos \theta /2\sin \theta ^{\prime }/2-e^{i(\varphi -\varphi ^{\prime
})}\sin \theta /2\cos \theta ^{\prime }/2,  \label{hu119}
\end{equation}

\begin{equation}
\phi ((-\frac 12)^{(\widehat{{\bf b}})};(+\frac 12)^{(\widehat{{\bf c}}%
)})=\sin \theta /2\cos \theta ^{\prime }/2-e^{i(\varphi -\varphi ^{\prime
})}\cos \theta /2\sin \theta ^{\prime }/2  \label{hu120}
\end{equation}
and 
\begin{equation}
\phi ((-\frac 12)^{(\widehat{{\bf b}})};(-\frac 12)^{(\widehat{{\bf c}}%
)})=\sin \theta /2\sin \theta ^{\prime }/2+e^{i(\varphi -\varphi ^{\prime
})}\cos \theta /2\cos \theta ^{\prime }/2.  \label{hu121}
\end{equation}
Hence, using Eqns. (\ref{si66}) - (\ref{si69}), we find that for the general
operator $[R]$ the elements are

\begin{eqnarray}
R_{11} &=&\left| \phi ((+\frac 12)^{(\widehat{{\bf b}})};(+\frac 12)^{(%
\widehat{{\bf c}})})\right| ^2r_1+\left| \phi ((+\frac 12)^{(\widehat{{\bf b}%
})};(-\frac 12)^{(\widehat{{\bf c}})})\right| ^2r_2  \nonumber \\
\ &=&[\cos ^2(\theta -\theta ^{\prime })/2-\sin \theta \sin \theta ^{\prime
}\sin ^2(\varphi -\varphi ^{\prime })/2]r_1  \nonumber \\
&&\ +[\sin ^2(\theta -\theta ^{\prime })/2+\sin \theta \sin \theta ^{\prime
}\sin ^2(\varphi -\varphi ^{\prime })/2]r_2  \label{hu122}
\end{eqnarray}

\begin{equation}
R_{12}=\left( \frac{r_1-r_2}2\right) (\sin \theta \cos \theta ^{\prime
}-\sin \theta ^{\prime }[\cos \theta \cos (\varphi -\varphi ^{\prime
})+i\sin (\varphi -\varphi ^{\prime })])  \label{hu123}
\end{equation}
\begin{equation}
R_{21}=\left( \frac{r_1-r_2}2\right) (\sin \theta \cos \theta ^{\prime
}-\sin \theta ^{\prime }[\cos \theta \cos (\varphi -\varphi ^{\prime
})-i\sin (\varphi -\varphi ^{\prime })])  \label{hu124}
\end{equation}
and 
\begin{eqnarray}
\ R_{22} &=&[\sin ^2(\theta -\theta ^{\prime })/2+\sin \theta \sin \theta
^{\prime }\sin ^2(\varphi -\varphi ^{\prime })/2]r_1  \nonumber \\
&&\ \ +[\cos ^2(\theta -\theta ^{\prime })/2-\sin \theta \sin \theta
^{\prime }\sin ^2(\varphi -\varphi ^{\prime })/2]r_2.  \label{hu125}
\end{eqnarray}

\subsection{{\bf Generalized Form of }$[\sigma _z]$ {\bf and} {\bf of its
Eigenvectors}}

Certain special operators are now readily obtained. If we seek the operator
for the projection of the spin along the new axis defined by $\widehat{{\bf c%
}}$, then we set $r_1=1$ and $r_2=-1.$ Hence the operator for the component
of spin in the direction $\widehat{{\bf c}}$ is

\begin{equation}
\lbrack \sigma _{\widehat{{\bf c}}}]=\left( 
\begin{array}{cc}
(\sigma _{\widehat{{\bf c}}})_{11} & (\sigma _{\widehat{{\bf c}}})_{12} \\ 
(\sigma _{\widehat{{\bf c}}})_{21} & (\sigma _{\widehat{{\bf c}}})_{22}
\end{array}
\right)  \label{hu126}
\end{equation}
where

\begin{eqnarray}
(\sigma _{\widehat{{\bf c}}})_{11} &=&\cos (\theta -\theta ^{\prime })-2\sin
\theta \sin \theta ^{\prime }\sin ^2(\varphi -\varphi ^{\prime })/2 
\nonumber \\
&=&\cos \theta \cos \theta ^{\prime }+\sin \theta \sin \theta ^{\prime }\cos
(\varphi -\varphi ^{\prime })  \label{hu127}
\end{eqnarray}
\begin{equation}
(\sigma _{\widehat{{\bf c}}})_{12}=\sin \theta \cos \theta ^{\prime }-\sin
\theta ^{\prime }[\cos \theta \cos (\varphi -\varphi ^{\prime })+i\sin
(\varphi -\varphi ^{\prime })]  \label{hu128}
\end{equation}
\begin{equation}
(\sigma _{\widehat{{\bf c}}})_{21}=\sin \theta \cos \theta ^{\prime }-\sin
\theta ^{\prime }[\cos \theta \cos (\varphi -\varphi ^{\prime })-i\sin
(\varphi -\varphi ^{\prime })]  \label{hu129}
\end{equation}
and 
\begin{equation}
(\sigma _{\widehat{{\bf c}}})_{22}=-\cos \theta \cos \theta ^{\prime }-\sin
\theta \sin \theta ^{\prime }\cos (\varphi -\varphi ^{\prime }).
\label{hu130}
\end{equation}

The operator for the square of the spin is obtained from the general formula
by setting $r_1=r_2=3.$ Hence

\begin{equation}
\lbrack \sigma ^2]=\left( 
\begin{array}{cc}
3 & 0 \\ 

0 & 3
\end{array}
\right) .  \label{hu131}
\end{equation}

The operator defined by the elements Eqns. (\ref{hu127}) - (\ref{hu130}) is
the most general for the component of the angular momentum along the axis of
quantization. We need the operators for the $x$ and $y$ components. We can
obtain these through the raising and lowering operators. But in order to
deduce these operators, we need the eigenvectors of the operator Eqn. (\ref
{hu126}), whose elements are given by Eqns. (\ref{hu127}) - (\ref{hu130}).
We could obtain these by direct calculation, but, given the form of the
operator, this is likely to be tedious. However, we can deduce these
eigenvectors by arguments based on the interpretation of the eigenvalue
equation (\ref{fi51}).

The elements of the eigenvectors are probability amplitudes. From the matrix
eigenvalue equation (\ref{fi51}), we see that their form is given by Eqns. (%
\ref{hu106}) - (\ref{hu109}). We only need to identify the initial and final
directions corresponding to these probability amplitudes. This we can do by
inspecting the five cases contained in Eqns. (\ref{si67}) - (\ref{ei82}).

The generalized operator $[\sigma _{\widehat{{\bf c}}}]$, Eqns. (\ref{hu127}%
) - (\ref{hu130}), is a function of only the two unit vectors $\widehat{{\bf %
b}}$ and $\widehat{{\bf c}}$ whose polar angles are $(\theta ,\varphi )$ and
($\theta ^{\prime },\varphi ^{\prime })$ respectively. Hence the elements of
its eigenvectors must also be functions of these four angles only. The
general case when the matrix and the eigenvectors are functions of two
vectors only is possible if $\widehat{{\bf a}}=\widehat{{\bf c}}$, but $%
\widehat{{\bf b}}\neq \widehat{{\bf a}}{\bf .}$ This corresponds to Case (d)
among the possibilities Eqns. (\ref{si69}) - (\ref{ei82}). The expectation
value of the quantity $R$ is in this case equal to the eigenvalue of $R$
corresponding to the initial state. This is true because $\widehat{{\bf c}}=%
\widehat{{\bf a}}$ means that the measurement is a repeat measurement. But
in order for the matrix expectation value to yield this result, it must be
that the vector is an eigenvector of the operator. We therefore conclude
that the eigenvectors contain as their elements probability amplitudes with
initial states corresponding to the unit vector $\widehat{{\bf a}}$ and
final states corresponding to the unit vector $\widehat{{\bf b}}$. As $%
\widehat{{\bf a}}=\widehat{{\bf c}}=(\sin \theta ^{\prime }\cos \varphi
^{\prime },\sin \theta ^{\prime }\sin \varphi ^{\prime },\cos \theta
^{\prime })$, and $\widehat{{\bf b}}=(\sin \theta \cos \varphi ,\sin \theta
\sin \varphi ,\cos \theta )$, we obtain the required eigenvectors as

\begin{equation}
\lbrack \xi _{\widehat{{\bf c}}}^{(+)}]=\left( 
\begin{array}{c}
\chi ((+\frac 12)^{(\widehat{{\bf a}})};(+\frac 12)^{(\widehat{{\bf b}})})
\\ 
\chi ((+\frac 12)^{(\widehat{{\bf a}})};(-\frac 12)^{(\widehat{{\bf b}})})
\end{array}
\right) =\left( 
\begin{array}{c}
\cos \theta ^{\prime }/2\cos \theta /2+e^{i(\varphi ^{\prime }-\varphi
)}\sin \theta ^{\prime }/2\sin \theta /2 \\ 
\cos \theta ^{\prime }/2\sin \theta /2-e^{i(\varphi ^{\prime }-\varphi
)}\sin \theta ^{\prime }/2\cos \theta /2
\end{array}
\right)  \label{hu132}
\end{equation}
and

\begin{equation}
\lbrack \xi _{\widehat{{\bf c}}}^{(-)}]=\left( 
\begin{array}{c}
\chi ((-\frac 12)^{(\widehat{{\bf a}})};(+\frac 12)^{(\widehat{{\bf b}})})
\\ 
\chi ((-\frac 12)^{(\widehat{{\bf a}})};(-\frac 12)^{(\widehat{{\bf b}})})
\end{array}
\right) =\left( 
\begin{array}{c}
\sin \theta ^{\prime }/2\cos \theta /2-e^{i(\varphi ^{\prime }-\varphi
)}\cos \theta ^{\prime }/2\sin \theta /2 \\ 
\sin \theta ^{\prime }/2\sin \theta /2+e^{i(\varphi ^{\prime }-\varphi
)}\cos \theta ^{\prime }/2\cos \theta /2
\end{array}
\right) .  \label{hu133}
\end{equation}

When we test these eigenvectors with their operator, we find that indeed the
eigenvalue equations 
\begin{equation}
\lbrack \sigma _{\widehat{{\bf c}}}][\xi _{\widehat{{\bf c}}%
}^{(+)}]=(+1)[\xi _{\widehat{{\bf c}}}^{(+)}]  \label{hu134}
\end{equation}
and 
\begin{equation}
\lbrack \sigma _{\widehat{{\bf c}}}][\xi _{\widehat{{\bf c}}%
}^{(-)}]=(-1)[\xi _{\widehat{{\bf c}}}^{(-)}]  \label{hu135}
\end{equation}
are satisfied.

\subsection{{\bf Generalized Forms of }$\left[ \sigma _x\right] ${\bf \ and }%
$[\sigma _y]$}

We are now able to deduce the generalized forms of the ladder operators and
hence of $\left[ \sigma _x\right] $ and $[\sigma _y]$. For this purpose we
use Eqns. (\ref{ei86}), (\ref{ei87}), (\ref{ei89}) and (\ref{ni90}) with the
eigenvectors of $[\sigma _{\widehat{{\bf c}}}]$, Eqns. (\ref{hu132}) and (%
\ref{hu133}). We find that the elements of $\left[ \sigma _{+}\right] $ are 
\begin{equation}
\left( \sigma _{+}\right) _{11}=-\sin \theta \cos \theta ^{\prime }\cos
(\varphi -\varphi ^{\prime })+\cos \theta \sin \theta ^{\prime }+i\sin
\theta \sin (\varphi ^{\prime }-\varphi )  \label{hu140}
\end{equation}

\begin{eqnarray}
\left( \sigma _{+}\right) _{12} &=&2[\cos ^2\theta /2\cos ^2\theta ^{\prime
}/2+\sin ^2\theta ^{\prime }/2\sin ^2\theta /2]\cos (\varphi -\varphi
^{\prime })+\sin \theta \sin \theta ^{\prime }  \nonumber \\
&&+2i[\sin ^2\theta /2\sin ^2\theta ^{\prime }/2-\cos ^2\theta ^{\prime
}/2\cos ^2\theta /2]\sin (\varphi ^{\prime }-\varphi )  \label{hu141}
\end{eqnarray}
\begin{eqnarray}
\left( \sigma _{+}\right) _{21} &=&-2[\sin ^2\theta /2\cos ^2\theta ^{\prime
}/2+\sin ^2\theta ^{\prime }/2\cos ^2\theta /2]\cos (\varphi -\varphi
^{\prime })+\sin \theta \sin \theta ^{\prime }  \nonumber \\
&&\ -2i[\sin ^2\theta ^{\prime }/2\cos ^2\theta /2-\sin ^2\theta /2\cos
^2\theta ^{\prime }/2]\sin (\varphi ^{\prime }-\varphi )  \label{hu142}
\end{eqnarray}
and 
\begin{equation}
\left( \sigma _{+}\right) _{22}=\sin \theta \cos \theta ^{\prime }\cos
(\varphi -\varphi ^{\prime })-\cos \theta \sin \theta ^{\prime }-i\sin
\theta \sin (\varphi ^{\prime }-\varphi ).  \label{hu143}
\end{equation}
The elements of $\left[ \sigma _{-}\right] $ are 
\begin{equation}
\left( \sigma _{-}\right) _{11}=-\sin \theta \cos \theta ^{\prime }\cos
(\varphi ^{\prime }-\varphi )+\cos \theta \sin \theta ^{\prime }-i\sin
\theta \sin (\varphi ^{\prime }-\varphi )  \label{hu144}
\end{equation}

\begin{eqnarray}
\left( \sigma _{-}\right) _{12} &=&-2[\cos ^2\theta /2\sin ^2\theta ^{\prime
}/2+\cos ^2\theta ^{\prime }/2\sin ^2\theta /2]\cos (\varphi ^{\prime
}-\varphi )+\sin \theta \sin \theta ^{\prime }  \nonumber \\
&&\ -2i[\cos ^2\theta ^{\prime }/2\sin ^2\theta /2-\cos ^2\theta /2\sin
^2\theta ^{\prime }/2]\sin (\varphi ^{\prime }-\varphi )  \label{hu145}
\end{eqnarray}

\begin{eqnarray}
\left( \sigma _{-}\right) _{21} &=&2[\sin ^2\theta /2\sin ^2\theta ^{\prime
}/2+\cos ^2\theta ^{\prime }/2\cos ^2\theta /2]\cos (\varphi ^{\prime
}-\varphi )+\sin \theta \sin \theta ^{\prime }  \nonumber \\
&&+2i[\cos ^2\theta /2\cos ^2\theta ^{\prime }/2-\sin ^2\theta ^{\prime
}/2\sin ^2\theta /2]\sin (\varphi ^{\prime }-\varphi )  \label{hu146} \\
&&\   \nonumber
\end{eqnarray}
and 
\begin{equation}
\left( \sigma _{-}\right) _{22}=\sin \theta \cos \theta ^{\prime }\cos
(\varphi ^{\prime }-\varphi )-\cos \theta \sin \theta ^{\prime }+i\sin
\theta \sin (\varphi ^{\prime }-\varphi ).  \label{hu147}
\end{equation}

Therefore the generalized operators for the $x$ and $y$ components of the
spin have the elements

\begin{equation}
\left( \sigma _x\right) _{11}=-\sin \theta \cos \theta ^{\prime }\cos
(\varphi ^{\prime }-\varphi )+\sin \theta ^{\prime }\cos \theta
\label{hu148}
\end{equation}
\begin{equation}
(\sigma _x)_{12}=\cos \theta \cos \theta ^{\prime }\cos (\varphi ^{\prime
}-\varphi )+\sin \theta \sin \theta ^{\prime }-i\cos \theta ^{\prime }\sin
(\varphi ^{\prime }-\varphi )  \label{hu149}
\end{equation}

\begin{equation}
(\sigma _x)_{21}=\cos \theta \cos \theta ^{\prime }\cos (\varphi ^{\prime
}-\varphi )+\sin \theta \sin \theta ^{\prime }+i\cos \theta ^{\prime }\sin
(\varphi ^{\prime }-\varphi )  \label{hu150}
\end{equation}

\begin{equation}
\left( \sigma _x\right) _{22}=\sin \theta \cos \theta ^{\prime }\cos
(\varphi ^{\prime }-\varphi )-\sin \theta ^{\prime }\cos \theta
\label{hu151}
\end{equation}

\begin{equation}
(\sigma _y)_{11}=\sin \theta \sin (\varphi ^{\prime }-\varphi )
\label{hu152}
\end{equation}
\begin{equation}
(\sigma _y)_{12}=-i\cos (\varphi ^{\prime }-\varphi )-\cos \theta \sin
(\varphi ^{\prime }-\varphi )  \label{hu153}
\end{equation}
\begin{equation}
(\sigma _y)_{21}=i\cos (\varphi ^{\prime }-\varphi )-\cos \theta \sin
(\varphi ^{\prime }-\varphi )  \label{hu154}
\end{equation}
and

\begin{equation}
(\sigma _y)_{22}=-\sin \theta \sin (\varphi ^{\prime }-\varphi ).
\label{hu155}
\end{equation}
We verify that

\begin{equation}
\left[ \sigma _x\right] ^2=[\sigma _y]^2=[\sigma _{\widehat{{\bf c}}}]^2=1.
\label{hu156}
\end{equation}
Also, the matrices anticommute:

\begin{equation}
\lbrack [\sigma _x],[\sigma _y]]_{+}=[[\sigma _y],[\sigma _{\widehat{{\bf c}}%
}]]_{+}=[[\sigma _{\widehat{{\bf c}}}],[\sigma _x]]_{+}=0.  \label{hu157}
\end{equation}
The commutators are

\begin{equation}
\lbrack [\sigma _x],[\sigma _y]]=2i[\sigma _{\widehat{{\bf c}}}]
\label{hu158}
\end{equation}

\begin{equation}
\lbrack [\sigma _y],[\sigma _{\widehat{{\bf c}}}]]=2i[\sigma _x]
\label{hu159}
\end{equation}
and 
\begin{equation}
\lbrack [\sigma _{\widehat{{\bf c}}}],[\sigma _x]]=2i[\sigma _y].
\label{hu160}
\end{equation}
Thus the generalized spin matrices obey the relations which the well-known
specialized forms, the Pauli spin matrices, Eqns. (\ref{ei83}), (\ref{ni92})
and (\ref{ni93}) obey.

At this point, it is worth observing that the $x$ and $y$ components of the
spin have been so identified purely because they are connected to the ladder
operators by the standard relations Eqns. (\ref{ei84}) and (\ref{ei85}). We
expect that in the limit in which the generalized form $[\sigma _{\widehat{%
{\bf c}}}]$ reduces to the Pauli spin matrix $[\sigma _z]$, these operators
will also reduce to the corresponding Pauli forms.

\subsection{\bf Special Forms of the Vectors and Operators }

{\bf \ }We need to verify that in the appropriate limits, the generalized
results obtained here reduce to the ones corresponding to the special cases.
Indeed, we find that we obtain the Pauli spin matrices by setting $\theta
=\theta ^{\prime }$ and $\varphi =\varphi ^{\prime }$.

We obtain the results Eqns. (\ref{ni95}) - (\ref{ni97}) by keeping$\;%
\widehat{{\bf a}}$ untouched, but setting $\theta =0$ and $\varphi =\pi $.
We then find that

\begin{equation}
\lbrack \sigma _{\widehat{{\bf c}}}]\rightarrow \left( 
\begin{array}{cc}
\cos \theta ^{\prime } & \sin \theta ^{\prime }e^{-i\varphi ^{\prime }} \\ 
\sin \theta ^{\prime }e^{i\varphi ^{\prime }} & -\cos \theta ^{\prime }
\end{array}
\right) ,  \label{hu161}
\end{equation}
\begin{equation}
\lbrack \xi _{\widehat{{\bf c}}}^{(+)}]\rightarrow \left( 
\begin{array}{c}
\cos \theta ^{\prime }/2 \\ 
-e^{i\varphi ^{\prime }}\sin \theta ^{\prime }/2
\end{array}
\right)  \label{hu162}
\end{equation}
and 
\begin{equation}
\lbrack \xi _{\widehat{{\bf c}}}^{(-)}]\rightarrow \left( 
\begin{array}{c}
\sin \theta ^{\prime }/2 \\ 
e^{i\varphi ^{\prime }}\cos \theta ^{\prime }/2
\end{array}
\right) .  \label{hu163}
\end{equation}

We see that we do indeed recover the operator Eqn. (\ref{ni95}) and the
eigenvectors Eqn. (\ref{ni96}) and Eqn. (\ref{ni97}). So the results given
by Eqns. (\ref{hu161}) - (\ref{hu163}) are consistent.

In the same limit, the ladder operators change as follows:

\begin{equation}
\left[ \sigma _{+}\right] \rightarrow \left( 
\begin{array}{cc}
\sin \theta ^{\prime } & -2\cos ^2\theta ^{\prime }/2e^{-i\varphi ^{\prime }}
\\ 
2\sin ^2\theta ^{\prime }/2e^{i\varphi ^{\prime }} & -\sin \theta ^{\prime }
\end{array}
\right)  \label{hu163a}
\end{equation}
and 
\begin{equation}
\left[ \sigma _{-}\right] \rightarrow \left( 
\begin{array}{cc}
\sin \theta ^{\prime } & 2\sin ^2\theta ^{\prime }/2e^{-i\varphi ^{\prime }}
\\ 
-2\cos ^2\theta ^{\prime }/2e^{i\varphi ^{\prime }} & -\sin \theta ^{\prime }
\end{array}
\right) .  \label{hu163b}
\end{equation}
Also 
\begin{equation}
\left[ \sigma _x\right] \rightarrow \left( 
\begin{array}{cc}
\sin \theta ^{\prime } & -e^{-i\varphi ^{\prime }}\cos \theta ^{\prime } \\ 
-e^{i\varphi ^{\prime }}\cos \theta ^{\prime } & -\sin \theta ^{\prime }
\end{array}
\right)  \label{hu163c}
\end{equation}
while 
\begin{equation}
\lbrack \sigma _y]\rightarrow \left( 
\begin{array}{cc}
0 & ie^{-i\varphi ^{\prime }} \\ 
-ie^{i\varphi ^{\prime }} & 0
\end{array}
\right) .  \label{hu163d}
\end{equation}

In order to obtain the Pauli spin matrices and vectors from the general
formulas we have had to set $\theta =\theta ^{\prime },\varphi =\varphi
^{\prime }.$ The operators in Eqns. (\ref{ni92}) and (\ref{ni93}) appear
from the general cases in the same limit as the Pauli operators and vectors.
Therefore in order to get them from the operators in Eqns. (\ref{hu163c})
and (\ref{hu163d}), we have to set $\theta ^{\prime }=0$, and $\varphi
^{\prime }=\pi ,$ since we have set $\theta =0$ and $\varphi =\pi $ to get
Eqns. (\ref{hu163c}) and (\ref{hu163d}). This is easily confirmed.

\subsection{\bf Connection with Standard Results }

We have now obtained the most general forms of the vectors and operators for
spin-$1/2$ systems. It is important to clarify the relation between these
quantities and the standard ones. It is of particular interest to expand on
how they are connected to the Pauli matrices and vectors.

The Pauli spin matrices Eqns. (\ref{ei83}), (\ref{ni92}) and (\ref{ni93})
are particular cases of the general cases Eqns. (\ref{hu127}) - (\ref{hu130}%
), Eqns. (\ref{hu148}) - (\ref{hu151}) and Eqns. (\ref{hu152}) - (\ref{hu155}%
). As we have seen, they arise from the generalized results when $\theta
=\theta ^{\prime }$ and $\varphi =\varphi ^{\prime }.$

Now, the generalized operators correspond to the probability amplitudes for
spin measurements having been expanded by the Land\'e formula using an
arbitrary vector to supply the intermediate states that appear in the
expansion. The less generalized formulas Eqns. (\ref{ni98}) - (\ref{hu101})
use the specialized vector $\widehat{{\bf d}}$ whose polar angles are $%
\theta =0$ and $\varphi =\pi $ to define the intermediate states. The Pauli
quantites result when the arbitrary vector used in the expansion equals the
direction of initial quantization.

The probability amplitudes which constitute the elements of the vectors
which comprise states in matrix mechanics are not just constants but have a
structure. This means that in the Pauli spin vectors 
\[
\lbrack \xi _{+}]=\left( 
\begin{array}{c}
1 \\ 
0
\end{array}
\right) \text{ and }[\xi _{-}]=\left( 
\begin{array}{c}
0 \\ 
1
\end{array}
\right) 
\]
the constants $1$ and $0$ which appear as the elements of the vectors are
probability amplitudes. When the probability amplitude is unity, it
represents a repeat measurement; when it is zero, it represents a
measurement to a state orthogonal to the initial state. Thus from $[\xi
_{+}] $ and $[\xi _{-}]$, we see that 
\[
\chi ((\pm \frac 12)^{(\widehat{{\bf a}})};(\pm \frac 12)^{(\widehat{{\bf a}}%
)})=1 
\]
while 
\[
\chi ((\pm \frac 12)^{(\widehat{{\bf a}})};(\mp \frac 12)^{(\widehat{{\bf a}}%
)})=0. 
\]
Here $\widehat{{\bf a}}$ is an arbitrary direction vector.

It is clear from Cases (a) - (c) enumerated above that the problem of
describing spin measurements can be approached in a variety of ways. The
task of computing the expectation value of a quantity depending on the spin
projection can be achieved in several ways. The most general way is
illustrated in Case (a), where the most general vectors and operators are
used. But the same results are obtained if we let the intermediate direction
equal the initial. Then the vectors simplify to the well-known ''up'' and
''down'' vectors, and the fact that the measurement is from one direction to
another is contained in the operator, which has its most general form.
Another way of simplifying the calculation is to let the intermediate
direction equal the final direction. In that case, the operator simplifies
to the Pauli matrix for the $z$ component of spin, while the vectors remain
general.

In the special case of computing the expectation value when the measurement
refers to the same initial and final direction, there are two possibilities.
The initial and the final directions are the same. If the intermediate
direction is not equal to the initial/final direction, then the vectors
remain general and the operator also remains general. This is the hard way
of doing it, because if the intermediate direction is chosen to equal the
initial/final direction, then the vectors reduce to the ''up'' and ''down''
vectors, and the operator to the Pauli matrix for the $z$ component of spin.

We observe that when we choose the intermediate direction to equal the
initial direction, as in Cases (b) and (e), this is formally equivalent to
expanding the probability amplitudes in the eigenrepresentation, because
this is when the operator matrices are diagonal.

\section{\bf \ SUMMARY AND DISCUSSION}

In this paper we have used the Land\'e interpretation of quantum mechanics
to investigate the connection between wave and matrix mechanics. We have
used Eqn. (\ref{nine}) to obtain a more detailed expression for the
expansion coefficients that appear in the usual expansion of a wave function
in terms of a basis set. We have reversed the expansion to obtain an
expansion of the basis functions in terms of the expansion coefficients and
the wave functions.

We have studied the transition from wave to matrix mechanics within the
framework of the Land\'e interpretation of quantum mechanics. By writing
down the matrix eigenvalue equation in a fully detailed way, we have
demonstrated that the elements of the eigenvectors resulting from the
solution of the eigenvalue equation are probability amplitudes with a
structure. These probability amplitudes describe a measurement starting from
the state characteristic of the eigenvalue of the matrix to one of the
states corresponding to the intermediate observable brought into operation
by expanding probability amplitudes by means of Eqn. (\ref{nine}). This
structure has been exploited to obtain the eigenvectors of complex matrices
by inspection.

In the derivation of the matrix treatment of orbital angular momentum, we
normally use the spherical harmonics as basis functions in order to obtain
the matrix angular momentum states and the matrix elements of the operators.
On the other hand, spin does not have a treatment proceeding from the
solution of a differential equation because. But we have shown that by using
the probability amplitudes for spin projection measurements, we can obtain
the matrix treatment of spin in the same way that we obtain the matrix forms
of the orbital angular momentum quantities. Thus, we have derived from first
principles the vectors and operators for spin $1/2$. With only the
assumption that the matrix operators and vectors thereby resulting have the
properties of angular momentum operators and states, we have derived
explicit expressions for the probability amplitudes. Moreover, we have
derived the most general forms for the spin operators. We have not seen
these forms in the literature; at any rate, this method of deriving these
quantities is new. In the appropriate limit, our operators and vectors
reduce to the Pauli spin vectors and matrices, which are thereby seen to be
special forms of the generalized results we have presented.

We make an interesting observation. A matrix-mechanics vector is
characteristic only of the eigenvalue corresponding to the state that
obtains before measurement, irrespective of the final state. Thus even zero
probability amplitudes may be represented by non-zero vectors. An example is
afforded by Eqns. (\ref{si69}) and (\ref{se70}). The matrix state is a
compact way of expressing discrete probability amplitudes, since a
particular vector represents all those probability amplitudes which belong
to the same initial state.

Apart from the new forms of the matrix quantities, we have devised a general
method for obtaining the generalized forms of spin operators and vectors. It
is natural to seek to extend this approach to the derivation of vectors and
matrix operators for other values of spin. We have successfully applied the
ideas presented here to the case $J=1.$ In this fashion, we have obtained
for $J=1$ quantities more generalized quantities than any so far seen in the
literature.

{\bf References}

1. Land\'e A, ''New Foundations of Quantum Mechanics'', Cambridge University
Press, 1965.

2. Land\'e A, ''From Dualism To Unity in Quantum Physics'', Cambridge
University Press, 1960.

3. See Ref. [1], p53.

\end{document}